\documentclass[epj]{svjour}
\usepackage{graphics}
\begin{document}
\title{A $SU(4)\otimes O(3)$ scheme for nonstrange baryons}
\author{P. Gonz\'alez\inst{1} \and J. Vijande\inst{1} \and A. Valcarce\inst{2}
\and H. Garcilazo\inst{3}}
\institute{
Dpto. de F\' \i sica Te\'orica and IFIC,
Universidad de Valencia - CSIC,
E-46100 Burjassot, Valencia, Spain \and
Grupo de F\'{\i}sica Nuclear and IUFFyM, Universidad de
Salamanca,
E-37008 Salamanca, Spain \and
Escuela Superior de F\'\i sica y Matem\'aticas,
Instituto Polit\'ecnico Nacional,
Edificio 9, 07738 M\'exico D.F., M\'exico}
\date{Received: date / Revised version: date}
%
\abstract{
We show that nonstrange baryon resonances can be classified according to
multiplets of $SU(4)\otimes O(3)$. We identify spectral regularities 
and degeneracies that allow us to predict the high spin
spectrum from 2 to 3 GeV.
} 
\maketitle
\section{Introduction}

The theoretical study of the high energy part of the baryonic spectrum has
been a subject of interest in the last decade, the aim being to get a better
understanding of the dynamics (in particular the confinement mechanism of
quarks in the baryon and the decay hadronization process), 
or, at least, the symmetries involved. In particular
the idea of a parity multiplet classification scheme at high excitation
energies as due to chiral symmetry was suggested some years ago \cite{Jid00}
and put in question later on \cite{Jaf05}. The lack of precise and complete
data prevents, at the current moment, to extract any definitive conclusion.

From the point of view of dynamics quark model refined potentials with linear
confinement have provided a reasonably accurate description of the whole
nonstrange light baryon spectrum once the coupling to $\pi N$ formation channels
is taken into account \cite{CapXX}. The low
probability obtained for many resonances to be formed would explain their no
experimental detection providing a solution to the so-called missing state
problem (the difference between the number of predicted states above 1 GeV
excitation energy -infinite with a linear potential- and the number of
known resonances).

Alternatively, the use of a quark-quark screened potential 
\cite{Vij04,Zha93,Gon06}, motivated by recent
unquenched QCD lattice calculations showing string breaking in the static
potential between two quarks \cite{Bal01,Bal05}, allows to obviate the
missing state problem (up to the limit of applicability of the model). In
Refs. \cite{Vij04} and \cite{Gon06} a correct prediction of the number and ordering of the
known $N$ and $\Delta $ resonances, up to 2.4 GeV mass or 1.5 GeV
excitation energy, is obtained. Above this limit the 3-free quark state is
energetically favored pointing out the need to implement the coupling to the
continuum. Nonetheless the unambiguous assignment of quantum numbers to
experimental states in the region of applicability translates, as we shall
show, into a well defined symmetry pattern.

In this article we identify the symmetry pattern as the one corresponding to 
$SU(4)\otimes O(3),$ $SU(4)$ containing $SU(2)_{\rm spin}\otimes SU(2)_{\rm isospin}$
and $O(3)$ standing for the orbital symmetry, and we analyze
spectral regularities and degeneracies according to it. 
The extension of this pattern to energies above the
applicability limit of the model allows us to predict the spectrum in
the range $2-3$ GeV where only incomplete and non-precise data exist.

\section{$SU(4)\otimes O(3)$ Pattern}

In Ref. \cite{Gon06} a quark model including confinement and minimal one
gluon exchange (coulomb + hyperfine) interactions has been developed.
Screening is imposed by requiring that the interaction potential saturates
(i.e., becomes constant) at a certain distance to be fixed
phenomenologically. Though one cannot obtain a precise fit to the spectrum
with such a simplistic model it is amazing that one can make
an unambiguous assignment of quantum numbers to the dominant
configuration of any $J^{P}$ ground and first non-radial states up to
$J=11/2$. This assignment agrees completely with the ones available in
the literature (only up to $J=7/2$) with much more refined
theoretical models \cite{CapYY} or purely phenomenological 
analysis \cite{Kle02}.

In Tables \ref{t1} and \ref{t2} we group
experimental resonances according to their dominant configuration (the
symmetry pattern obtained has been extended up to 2 GeV excitation
energy). To express the spatial part we use the hyperspherical harmonic
notation, i.e., the quantum numbers $(K,L,Symmetry)$. The so-called great
orbital, $K,$ defines the parity of the state, $P=(-)^{K}$, and its
centrifugal barrier energy, $\frac{{\cal L}({\cal L}+1)}{2m\left\langle \rho
^{2}\right\rangle }$ (${\cal L}=K+\frac{3}{2}$, $\rho $: hyperrradius). $L$
is the total orbital angular momentum. $Symmetry$ specifies the spatial
symmetry $(\left[ 3\right] :$ symmetric, $\left[ 21\right] :$ mixed, $\left[
111\right] :$ antisymmetric$)$ which combines to the spin, $S,$ and isospin, 
$T,$ symmetries $(S,T=3/2:$ symmetric; $S,T=1/2:$ mixed) to have a symmetric
wave function (the color part is antisymmetric). More precisely, $T=1/2$ for 
$N$ and $T=3/2$ for $\Delta $, hence the spatial-spin wave function must be
mixed for $N$ and symmetric for $\Delta $.

\begin{table*}[h!!]
\caption{Positive parity $N$ and $\Delta $ states (masses in MeV) for
different dominant spatial-spin configurations up to $\simeq 3$ GeV.
Experimental data are from PDG \protect\cite{Eid04}. Stars have been omitted
for four-star resonances. States denoted by a question mark correspond to
predicted resonances that do not appear in the PDG (their predicted masses
appear in Table \protect\ref{t3}).}
\begin{center}
\begin{tabular}{|c|c|c|} \hline
$(K,L,Symmetry)$ & $S=1/2$ & $S=3/2$ \\ \hline
$(0,0,[3])$ & $N(1/2^{+})(940)$ & \\ \cline{2-3}
& & $\Delta (3/2^{+})(1232)$ \\ \hline
$(2,2,[3])$ & $N(5/2^{+})(1680),N(3/2^{+})(1720)$ & \\ \cline{2-3}
& & $\Delta (7/2^{+})(1950)$ \\ \hline
$(4,4,[3])$ & $N(9/2^{+})(2220)$ & \\ \cline{2-3}
& & $\Delta (11/2^{+})(2420)$ \\ \hline
$(6,6,[3])$ & $N(13/2^{+})(\ast \ast )(2700)$ & \\ \cline{2-3}
& & $\Delta (15/2^{+})(\ast \ast )(2950)$ \\ \hline\hline
             & $N(1/2^{+})(\ast \ast \ast )(1710)$ & \\ \cline{2-3}
$(2,0,[21])$ &  & \\ \cline{2-3}
& $\Delta (1/2^{+})(1750)$ & \\ \hline
& $N(5/2^{+})(\ast \ast )(2000)$ & 
 \\ \cline{2-3}
$(2,2,[21])$ &  & $N(7/2^{+})(\ast \ast )(1990)
$ \\ \cline{2-3}
& $\Delta (5/2^{+})(1905)$ & \\ \hline
 & $N(9/2^{+})(2220)$ &  \\ \cline{2-3}
$(4,4,[21])$ &  & $N(11/2^{+})(?)$ \\ \cline{2-3}
& $\Delta (9/2^{+})(\ast \ast )(2300)$ &\\ \hline
 & $N(13/2^{+})(2700)$ &  \\ \cline{2-3}
$(6,6,[21])$ &  & $N(15/2^{+})(?)$ \\ \cline{2-3}
& $\Delta (13/2^{+})(?)$ & \\ \hline
\end{tabular}
\label{t1}
\end{center}
\end{table*}

\begin{table*}[h!!]
\caption{Negative parity $N$ and $\Delta $ states (masses in MeV) for
different dominant spatial-spin configurations up to $\simeq 3$ GeV.
Experimental data are from PDG \protect\cite{Eid04}. Stars have been omitted
for four-star resonances. States denoted by a question mark correspond to
predicted resonances that do not appear in the PDG (their predicted masses
appear in Table \protect\ref{t3}).}
\begin{center}
\begin{tabular}{|c|c|c|} \hline
$(K,L,Symmetry)$ & $S=1/2$ & $S=3/2$ \\ \hline
& $N(3/2^{-})(1520),N(1/2^{-})(1535)$ & \\ \cline{2-3}
$(1,1,[21])$ & & $N(5/2^{-})(1675)$ \\ \cline{2-3}
& $\Delta (3/2^{-})(1700),\Delta (1/2^{-})(1620)$ & \\ \hline
& $N(7/2^{-})(2190)$ & \\ \cline{2-3}
$(3,3,[21])$ & & $N(9/2^{-})(2250)$ \\ \cline{2-3}
& $\Delta (7/2^{-})(\ast )(2200)$ & \\ \hline
& $N(11/2^{-})(\ast \ast \ast )(2600)$ & \\ \cline{2-3}
$(5,5,[21])$ & & $N(13/2^{-})(?)$ \\ \cline{2-3}
& $\Delta (11/2^{-})(?)$ & \\ \hline\hline
$(3,3,[3])$ & $N(7/2^{-})(?)$ & \\ \cline{2-3}
&  & $\Delta (9/2^{-})(\ast \ast )(2400)$ \\ 
\hline
$(5,5,[3])$ & $N(11/2^{-})(?)$ & \\ \cline{2-3}
 &  & $\Delta (13/2^{-})(\ast \ast )(2750)$
\\ \hline
\end{tabular}
\end{center}
\label{t2}
\end{table*}

A look at the Tables makes manifest the underlying symmetry.

For positive parity each box in the upper part of Table \ref{t1} groups four 
$N(J^{P})$ ground states -two isospin and two spin projections- and sixteen 
$\Delta (J^{P})$ ground states -four isospin and four spin projections-
corresponding to an orbitally symmetric configuration. It is worth to
mention that the $N$ states have also some probability of mixed\ orbital
symmetry with the same values of $K$ and $L.$ This mixing explains the
appearance of an additional $N(3/2^+),$ which is the symmetry
partner of $\Delta (1/2^+),$ in the second upper box as a
consequence of the bigger hyperfine attraction for orbitally symmetric
$S=1/2$ states. Another consequence of the mixing is the presence 
of corresponding $N$
excitations with reverse probabilities that appear in the boxes of the lower
part of Table I. Any of these excitations adds four states -two isospin and
two spin projections- to the eight $N$'s and eight $\Delta $'s ground states
present in each box.

For negative parity, Table \ref{t2}, the same box pattern repeats with different
combinations of $N$'s and $\Delta $'s. 

This 20 member box picture where all the members of the same box have the
same parity given by $P=(-)^{L}$, is a reflection of an underlying 
$SU(4)\otimes O(3)$ symmetry providing a $(20,L^{P})$ classification scheme,
the 20plet structure coming out naturally from the product of irreducible
quark representations: $4\otimes 4\otimes 4=20_{S}\oplus 20_{M}\oplus 20_{M}
\oplus \overline{4}$.
The only difference in content between a box and the corresponding 20plet
refers to mixed $N$ resonances being a linear combination of $N$ members
of the 20plets with well defined orbital symmetry.

It is worth to emphasize that $SU(4)$ goes beyond a factorization 
$SU(2) \otimes SU(2)$ as can be checked 
through the $N-\Delta$ degeneracies appearing within the same box when the 
$SU(4)$ breaking spin-spin interaction plays a minor role.

\section{Spectral regularities and degeneracies}

From the spectral pattern represented by Tables \ref{t1} and \ref{t2} experimental
regularities and degeneracies for $J\geq 5/2$ ground states come out:

\bigskip

i) $E_{N,\Delta }(J+2)-E_{N,\Delta }(J)\approx 400-500$ MeV

\smallskip
\smallskip

ii) $N(J^{\pm})\approx \Delta (J^{\pm})$ for $J=\frac{4n+3}{2}$, $n=1,2...$

\smallskip
\smallskip

iii) $N(J^{+})\approx N(J^{-})$ for $J=\frac{4n+1}{2}$, $n=1,2.$

\bigskip

\noindent
These rules can also 
be obtained theoretically by refitting the quark model to reproduce 
precisely the $J \ge 5/2$ states. Rule i) expresses the
increasing of the centrifugal barrier between states with the same orbital
symmetry and the slowly varying spin-spin contribution for the 
same $S$ when increasing $L$ for $L \geq 2$ . 
Rule ii) for positive parity reflects the small spin-spin contribution
for $S=3/2$ when $L \geq 2$, and for negative parity reflects the $SU(4)\otimes O(3)$
degeneracy for $N^{\prime }s$ and $\Delta ^{\prime }s$ in the same multiplet
once the centrifugal barrier suppresses greatly the hyperfine splitting.
Rule iii) for $N$ parity doublets comes from the balance between a bigger
repulsion (due to bigger $K$ and $L$) and a bigger hyperfine attraction (due
to lower $S$) for $N(J^{+})$ against $N(J^{-})$. No parallel degeneracy
for $\Delta$'s is found since the $J^-$ states should be higher in 
mass (bigger centrifugal repulsion and spin-spin repulsion as well)
than the $J^+$ ones. These results are in disagreement with 
the parity multiplet classification scheme proposed in Ref. \cite{Jid00}.

For excited states the absence of spin-orbit and tensor forces in our
dynamical model suggests a new rule for $J\geq 5/2$:

\bigskip

iv) $(N(J),\Delta (J))^{^{\bullet }}\approx (N(J+1),\Delta (J+1))$

\bigskip
\noindent
say the first non-radial excitation of $N(J)$ and the ground state of $N(J+1)$
are almost degenerate (the same for $\Delta$).
This rule is well satisfied by experimental data.

Taking into account rules i)-iv) and the developed symmetry pattern we can
make predictions for, until now, unknown states from 2 to 3 GeV, Table \ref{t3}. 
Though some of our predicted states might be masked by experimental
uncertainties and others could not be easily detected (small coupling to
formation channels) we hope the results in Table \ref{t3} may be of 
some help to guide future experimental searches.

\begin{table*}[tbp]
\caption{Predicted $N$ and $\Delta $ states in the interval $[2.2,3.0]$ MeV.
We denote by a black dot the first non-radial excitation.}
\begin{center}
\begin{tabular}{|c|c|c||c|c|} \hline
& \multicolumn{2}{c||}{$N$} & \multicolumn{2}{c|}{$\Delta$} \\ \hline
$J=7/2$ & $N(7/2^+)^\bullet$(2220) & $N(7/2^-)^\bullet$(2250) & & $
\Delta(7/2^-)^\bullet$(2400) \\ 
$J=9/2$ & $N(9/2^+)^\bullet$(2450) & $N(9/2^-)^\bullet$(2600) & $
\Delta(9/2^+)^\bullet$(2420) & $\Delta(9/2^-)^\bullet$(2650) \\ 
$J=11/2$ & $N(11/2^+)$(2450) & & & $\Delta(11/2^-)$(2650) \\ 
& $N(11/2^+)^\bullet$(2700) & $N(11/2^-)^\bullet$(2650) & $
\Delta(11/2^+)^\bullet$(2850) & $\Delta(11/2^-)^\bullet$(2750) \\ 
$J=13/2$ & & $N(13/2^-)$(2650) & $\Delta(13/2^+)$(2850) & \\ 
& $N(13/2^+)^\bullet$(2900) & & $\Delta(13/2^+)^\bullet$(2950) & \\ 
$J=15/2$ & $N(15/2^+)$(2900) & & & 
\\ \hline
\end{tabular}
\end{center}
\label{t3}
\end{table*}

\bigskip

This work has been partially funded by MCyT
under Contract No. FPA2004-05616, by JCyL under
Contract No. SA104/04, and by GV under Contract No.
GV05/276.

\end{document}